\documentclass[conference]{IEEEtran}
\pdfoutput=1
\usepackage[utf8]{inputenc}
\usepackage{fst}
\usepackage{authblk}

\begin{document}

\title{PAM-6 Coded Modulation for IM/DD Channels with a Peak-Power Constraint}

\author{Tobias~Prinz\IEEEauthorrefmark{1}, Thomas~Wiegart\IEEEauthorrefmark{1}, Daniel~Plabst\IEEEauthorrefmark{1},\\Stefano~Calabr\`{o}\IEEEauthorrefmark{2}, Georg~B\"ocherer\IEEEauthorrefmark{2}, Nebojsa~Stojanovic\IEEEauthorrefmark{2}, Talha~Rahman\IEEEauthorrefmark{2}\\
\IEEEauthorrefmark{1}Institute for Communications Engineering, Technical University of Munich\\
\IEEEauthorrefmark{2}German Research Center, Huawei Technologies Duesseldorf GmbH\\
Email: \{tobias.prinz, thomas.wiegart, d.plabst\}@tum.de\\
\{stefano.calabro, georg.bocherer, nebojsa.stojanovic, talha.rahman\}@huawei.com}

\maketitle

\begin{abstract}
Coded modulation with probabilistic amplitude shaping (PAS) is considered for intensity modulation/direct detection channels with a transmitter peak-power constraint. PAS is used to map bits to a uniform PAM-6 distribution and outperforms PAM-8 for rates up to around 2.3 bits per channel use.
PAM-6 with PAS also outperforms a cross-shaped QAM-32 constellation by up to \SI{1}{dB} and \SI{0.65}{dB} after bit-metric soft- and hard decoding, respectively. An alternative PAM-6 scheme based on a framed-cross-shaped QAM-32 constellation is proposed that shows similar gains.
\end{abstract}

\section{Introduction}
Short-reach optical links are often based on transceivers that use intensity modulation (IM) and direct detection (DD). This allows for cheap hardware, low power consumption, and low latency \cite{chagnon_2019_optical, zhong_2018_digital}.
Since short-reach systems are usually operated without optical amplifiers, practical constraints at the transmitter laser and modulator dictate a peak power constraint~ \cite{wiegart_2020_probabilistically}.

Early IM/DD systems were based on on-off keying (OOK), while newer systems use unipolar \mbox{4-ary} and \mbox{8-ary} pulse amplitude modulation (PAM) formats to increase data rates. The higher-order modulation comes at the price of increased complexity, especially at the receiver.
The number of symbols in a modulation alphabet is usually a power of two to allow for 
an easy integration with binary data and binary forward error correction (FEC). However, under a peak power constraint, other modulation alphabets give larger information rates, especially for small alphabet sizes. 

We study three schemes to transmit \mbox{PAM-6}. First, we review the benchmark cross-shaped \mbox{QAM-32} constellation shown in Fig.~\ref{fig:qam32_corner} below~\cite{ghiasi_2012_investigation, chorchos_2019_pam6}. Second, we propose a modified \mbox{QAM-32} constellation with a framed-cross shape which improves information rates. Third, we use probabilistic amplitude shaping (PAS)~\cite{boecherer2015pas} to map bits to uniformly-spaced signal constellations whose number of points~$M$ is even but not a power of two. The idea is that a distribution matcher (DM) creates a block of \mbox{PAM-$(M/2)$} symbols that are combined with parity bits from a FEC code to obtain a block of \mbox{PAM-$M$} symbols. 
Of course, PAS algorithms can also put out non-uniform distributions.

\begin{figure*}
    \centering
    \subfloat[SEs with soft SMD.]{
        \footnotesize
        \begin{tikzpicture} 
\footnotesize
\begin{axis}[
xlabel={PSNR [\si{dB}]},
width=.48\textwidth, 
height=6cm,
ylabel={$\Rsmd$ (bpcu)},
grid=both,
legend cell align={left},
 legend style ={at={(0,1)},anchor=north west, fill opacity=0.5, text opacity=1},
ymin=1.8,
ymax=2.2,
xmin=20.5,
xmax=24
]

\addplot[name path global=pam4,line width=1,black] table[x=psnr,y=rate] {results/rates_4pam_smd_uniform.txt};
\addlegendentry{\scriptsize PAM-4};

\addplot[name path global=pam8,line width=1,black,dashed] table[x=psnr,y=rate] {results/rates_8pam_smd_uniform.txt};
\addlegendentry{\scriptsize PAM-8};

\addplot[name path global=pam6,line width=1.0,TUMBeamerRed] table[x=psnr,y=rate] {results/rates_6pam_smd_uniform.txt};
\addlegendentry{\scriptsize PAM-6};

\addplot[name path global=pam6shaped,line width=1,TUMBeamerRed,dotted] table[x=psnr,y=rate] {results/rates_6pam_smd_shaped.txt};
\addlegendentry{\scriptsize PAM-6 shaped};

\addplot[name path global=qam32cross,line width=1.5,TUMBeamerGreen] table[x=psnr,y=rate] {results/rates_cross_32qam_smd_uniform.txt};
\addlegendentry{\scriptsize Cross QAM-32/PAM-6};
\addplot[name path global=qam32grid,line width=1.5,TUMBeamerOrange] table[x=psnr,y=rate] {results/rates_grid_32qam_smd_uniform.txt};
\addlegendentry{\scriptsize opt QAM-32/PAM-6};
\addplot[name path global=qam32framedcross,line width=1.5,TUMBeamerBlue,dashed] table[x=psnr,y=rate] {results/rates_framed_cross_32qam_smd_uniform.txt};
\addlegendentry{\scriptsize Fr. Cross QAM-32/PAM-6};

\path[name path global=line] (axis cs:\pgfkeysvalueof{/pgfplots/xmin},1.847) -- (axis cs: \pgfkeysvalueof{/pgfplots/xmax},1.847);
\path[name intersections={of=line and pam6, name=p1}, name intersections={of=line and pam8, name=p2}];
\draw[<->,thick] let \p1=(p1-1), \p2=(p2-1) in (p1-1) -- ([xshift=-0.05cm]p2-1) node [below,yshift=-0.1cm,midway,fill=white, fill opacity=0.7,text opacity=1, draw opacity=1] {%
        \pgfplotsconvertunittocoordinate{x}{\x1}%
        \pgfplotscoordmath{x}{datascaletrafo inverse to fixed}{\pgfmathresult}%
        \edef\valueA{\pgfmathresult}%
        \pgfplotsconvertunittocoordinate{x}{\x2}%
        \pgfplotscoordmath{x}{datascaletrafo inverse to fixed}{\pgfmathresult}%
        \pgfmathparse{\pgfmathresult - \valueA}%
        \pgfmathprintnumber{\pgfmathresult} dB
};

\path[name path global=line] (axis cs:\pgfkeysvalueof{/pgfplots/xmin},1.853) -- (axis cs: \pgfkeysvalueof{/pgfplots/xmax},1.853);
\path[name intersections={of=line and pam6shaped, name=p1}, name intersections={of=line and pam8, name=p2}];
\draw[<->,thick] let \p1=(p1-1), \p2=(p2-1) in (p1-1) -- ([xshift=-0.05cm]p2-1) node [above,yshift=0.1cm,midway,fill=white, fill opacity=0.7,text opacity=1, draw opacity=1] {%
        \pgfplotsconvertunittocoordinate{x}{\x1}%
        \pgfplotscoordmath{x}{datascaletrafo inverse to fixed}{\pgfmathresult}%
        \edef\valueA{\pgfmathresult}%
        \pgfplotsconvertunittocoordinate{x}{\x2}%
        \pgfplotscoordmath{x}{datascaletrafo inverse to fixed}{\pgfmathresult}%
        \pgfmathparse{\pgfmathresult - \valueA}%
        \pgfmathprintnumber{\pgfmathresult} dB
};

\end{axis}

\end{tikzpicture}
        \label{fig:rates_smd_soft}

    }
    \hfill
    \subfloat[SEs with soft BMD.]{
        \footnotesize
        \begin{tikzpicture}
\footnotesize
\begin{axis}[
xlabel={PSNR [\si{dB}]},
width=.48\textwidth, 
height=6cm,
ylabel={$\Rbmd$ (bpcu)},
grid=both,
legend cell align={left},
 legend style ={at={(0,1.0)},anchor=north west, fill opacity=0.5, text opacity=1},
ymin=1.7,
ymax=2.3,
xmin=20,
xmax=24
]

\addplot[name path global=pam4,line width=1,black] table[x=psnr,y=rate] {results/rates_4pam_bmd_uniform.txt};
\addlegendentry{\scriptsize PAM-4};

\addplot[name path global=pam8,line width=1,black, dashed] table[x=psnr,y=rate] {results/rates_8pam_bmd_uniform.txt};
\addlegendentry{\scriptsize PAM-8};

\addplot[name path global=pam6,line width=1,TUMBeamerRed] table[x=psnr,y=rate] {results/rates_6pam_bmd_uniform.txt};
\addlegendentry{\scriptsize PAM-6};

\addplot[name path global=pam6shaped,line width=1,TUMBeamerRed, dotted] table[x=psnr,y=rate] {results/rates_6pam_bmd_shaped.txt};
\addlegendentry{\scriptsize PAM-6 shaped};

\addplot[name path global=qam32cross,line width=1.5,TUMBeamerGreen] table[x=psnr,y=rate] {results/rates_cross_32qam_bmd_uniform.txt};
\addlegendentry{\scriptsize CrossQAM-32/PAM-6};

\addplot[name path global=qam32framedcross,line width=1.5,TUMBeamerBlue,dashed] table[x=psnr,y=rate] {results/rates_framed_cross_32qam_bmd_uniform.txt};
\addlegendentry{\scriptsize Fr. Cross QAM-32/PAM-6};

\path[name path global=line] (axis cs:\pgfkeysvalueof{/pgfplots/xmin},1.865) -- (axis cs: \pgfkeysvalueof{/pgfplots/xmax},1.865);
\path[name intersections={of=line and pam6, name=p1}, name intersections={of=line and pam8, name=p2}];
\draw[<->,thick] let \p1=(p1-1), \p2=(p2-1) in (p1-1) -- ([xshift=-0.05cm]p2-1) node [below,yshift=-0.1cm,midway,fill=white, fill opacity=0.7,text opacity=1, draw opacity=1] {%
        \pgfplotsconvertunittocoordinate{x}{\x1}%
        \pgfplotscoordmath{x}{datascaletrafo inverse to fixed}{\pgfmathresult}%
        \edef\valueA{\pgfmathresult}%
        \pgfplotsconvertunittocoordinate{x}{\x2}%
        \pgfplotscoordmath{x}{datascaletrafo inverse to fixed}{\pgfmathresult}%
        \pgfmathparse{\pgfmathresult - \valueA}%
        \pgfmathprintnumber{\pgfmathresult} dB
};

\path[name path global=line] (axis cs:\pgfkeysvalueof{/pgfplots/xmin},2.01) -- (axis cs: \pgfkeysvalueof{/pgfplots/xmax},2.01);
\path[name intersections={of=line and pam6, name=p1}, name intersections={of=line and pam8, name=p2}];
\draw[<->,thick] let \p1=(p1-1), \p2=(p2-1) in (p1-1) -- ([xshift=-0.05cm]p2-1) node [above,yshift=0.1cm,midway,fill=white, fill opacity=0.7,text opacity=1, draw opacity=1] {%
        \pgfplotsconvertunittocoordinate{x}{\x1}%
        \pgfplotscoordmath{x}{datascaletrafo inverse to fixed}{\pgfmathresult}%
        \edef\valueA{\pgfmathresult}%
        \pgfplotsconvertunittocoordinate{x}{\x2}%
        \pgfplotscoordmath{x}{datascaletrafo inverse to fixed}{\pgfmathresult}%
        \pgfmathparse{\pgfmathresult - \valueA}%
        \pgfmathprintnumber{\pgfmathresult} dB
};

\path[name path global=line] (axis cs:\pgfkeysvalueof{/pgfplots/xmin},1.99) -- (axis cs: \pgfkeysvalueof{/pgfplots/xmax},1.99);
\path[name intersections={of=line and pam6, name=p1}, name intersections={of=line and qam32cross, name=p2}];
\draw[<->,thick] let \p1=(p1-1), \p2=(p2-1) in (p1-1) -- ([xshift=-0.05cm]p2-1) node [below,yshift=-0.1cm,midway,fill=white, fill opacity=0.7,text opacity=1, draw opacity=1] {%
        \pgfplotsconvertunittocoordinate{x}{\x1}%
        \pgfplotscoordmath{x}{datascaletrafo inverse to fixed}{\pgfmathresult}%
        \edef\valueA{\pgfmathresult}%
        \pgfplotsconvertunittocoordinate{x}{\x2}%
        \pgfplotscoordmath{x}{datascaletrafo inverse to fixed}{\pgfmathresult}%
        \pgfmathparse{\pgfmathresult - \valueA}%
        \pgfmathprintnumber{\pgfmathresult} dB
};

\path[name path global=line] (axis cs:\pgfkeysvalueof{/pgfplots/xmin},1.875) -- (axis cs: \pgfkeysvalueof{/pgfplots/xmax},1.875);
\path[name intersections={of=line and pam6shaped, name=p1}, name intersections={of=line and pam8, name=p2}];
\draw[<->,thick] let \p1=(p1-1), \p2=(p2-1) in (p1-1) -- ([xshift=-0.05cm]p2-1) node [above,yshift=0.1cm,midway,fill=white, fill opacity=0.7,text opacity=1, draw opacity=1] {%
        \pgfplotsconvertunittocoordinate{x}{\x1}%
        \pgfplotscoordmath{x}{datascaletrafo inverse to fixed}{\pgfmathresult}%
        \edef\valueA{\pgfmathresult}%
        \pgfplotsconvertunittocoordinate{x}{\x2}%
        \pgfplotscoordmath{x}{datascaletrafo inverse to fixed}{\pgfmathresult}%
        \pgfmathparse{\pgfmathresult - \valueA}%
        \pgfmathprintnumber{\pgfmathresult} dB
};

\end{axis}

\end{tikzpicture}
        \label{fig:rates_bmd_soft}

    }\\
    \subfloat[SEs with SMD using hard decisions.]{
        \footnotesize
        \begin{tikzpicture} 
\footnotesize
\begin{axis}[
xlabel={PSNR [\si{dB}]},
width=.48\textwidth, 
height=5.5cm,
ylabel={$\Rhdsmd$ (bpcu)},
grid=both,
legend cell align={left},
 legend style ={at={(0,1.0)},anchor=north west, fill opacity=0.5, text opacity=1},
ymin=1.8,
ymax=2.4,
xmin=23.5,
xmax=27
]

\addplot[name path global=pam4,line width=1,black] table[x=psnr,y=rate] {results/rates_4pam_smd_hard_uniform.txt};
\addlegendentry{\scriptsize PAM-4};

\addplot[name path global=pam8,line width=1,black,dashed] table[x=psnr,y=rate] {results/rates_8pam_smd_hard_uniform.txt};
\addlegendentry{\scriptsize PAM-8};

\addplot[name path global=pam6,line width=1,TUMBeamerRed] table[x=psnr,y=rate] {results/rates_6pam_smd_hard_uniform.txt};
\addlegendentry{\scriptsize PAM-6};

\addplot[name path global=qam32cross,line width=1.5,TUMBeamerGreen] table[x=psnr,y=rate] {results/rates_cross_32qam_smd_hard.txt};
\addlegendentry{\scriptsize Cross QAM-32/PAM-6};

\addplot[name path global=qam32gird,line width=1.5,TUMBeamerOrange] table[x=psnr,y=rate] {results/rates_grid_32qam_smd_hard.txt};
\addlegendentry{\scriptsize opt QAM-32/PAM-6};

\addplot[name path global=qam32framedcross,line width=1.5,TUMBeamerBlue,dashed] table[x=psnr,y=rate] {results/rates_framed_cross_32qam_smd_hard.txt};
\addlegendentry{\scriptsize Fr. Cross QAM-32/PAM-6};

\path[name path global=line] (axis cs:\pgfkeysvalueof{/pgfplots/xmin},1.99) -- (axis cs: \pgfkeysvalueof{/pgfplots/xmax},1.99);
\path[name intersections={of=line and pam6, name=p1}, name intersections={of=line and pam8, name=p2}];
\draw[<->,thick] let \p1=(p1-1), \p2=(p2-1) in (p1-1) -- ([xshift=-0.05cm]p2-1) node [below,yshift=-0.1cm,midway,fill=white, fill opacity=0.7,text opacity=1, draw opacity=1] {%
        \pgfplotsconvertunittocoordinate{x}{\x1}%
        \pgfplotscoordmath{x}{datascaletrafo inverse to fixed}{\pgfmathresult}%
        \edef\valueA{\pgfmathresult}%
        \pgfplotsconvertunittocoordinate{x}{\x2}%
        \pgfplotscoordmath{x}{datascaletrafo inverse to fixed}{\pgfmathresult}%
        \pgfmathparse{\pgfmathresult - \valueA}%
        \pgfmathprintnumber{\pgfmathresult} dB
};

\end{axis}

\end{tikzpicture}
        \label{fig:rates_smd_hard}
    }
    \hfill
    \subfloat[SEs with BMD using hard decisions.]{
        \footnotesize
        \begin{tikzpicture} 
\footnotesize
\begin{axis}[
xlabel={PSNR [\si{dB}]},
width=.48\textwidth, 
height=5.5cm,
ylabel={$\Rhdbmd$ (bpcu)},
grid=both,
legend cell align={left},
 legend style ={at={(0,1.0)},anchor=north west, fill opacity=0.5, text opacity=1},
ymin=1.8,
ymax=2.4,
xmin=23,
xmax=27
]

\addplot[name path global=pam4,line width=1,black] table[x=psnr,y=rate] {results/rates_4pam_bmd_hard_uniform.txt};
\addlegendentry{\scriptsize PAM-4};

\addplot[name path global=pam8,line width=1,black,dashed] table[x=psnr,y=rate] {results/rates_8pam_bmd_hard_uniform.txt};
\addlegendentry{\scriptsize PAM-8};

\addplot[name path global=pam6,line width=1,TUMBeamerRed] table[x=psnr,y=rate] {results/rates_6pam_bmd_hard_uniform.txt};
\addlegendentry{\scriptsize PAM-6};

\addplot[name path global=qam32cross,line width=1.5,TUMBeamerGreen] table[x=psnr,y=rate] {results/rates_cross_32qam_bmd_hard.txt};
\addlegendentry{\scriptsize Cross QAM-32/PAM-6};

\addplot[name path global=qam32framedcross,line width=1.5,TUMBeamerBlue,dashed] table[x=psnr,y=rate] {results/rates_framed_cross_32qam_bmd_hard.txt};
\addlegendentry{\scriptsize Fr. Cross QAM-32/PAM-6};

\path[name path global=line] (axis cs:\pgfkeysvalueof{/pgfplots/xmin},2) -- (axis cs: \pgfkeysvalueof{/pgfplots/xmax},2);
\path[name intersections={of=line and pam6, name=p1}, name intersections={of=line and pam8, name=p2}];
\draw[<->,thick] let \p1=(p1-1), \p2=(p2-1) in (p1-1) -- ([xshift=-0.05cm]p2-1) node [below,yshift=-0.1cm,midway,fill=white, fill opacity=0.7,text opacity=1, draw opacity=1] {%
        \pgfplotsconvertunittocoordinate{x}{\x1}%
        \pgfplotscoordmath{x}{datascaletrafo inverse to fixed}{\pgfmathresult}%
        \edef\valueA{\pgfmathresult}%
        \pgfplotsconvertunittocoordinate{x}{\x2}%
        \pgfplotscoordmath{x}{datascaletrafo inverse to fixed}{\pgfmathresult}%
        \pgfmathparse{\pgfmathresult - \valueA}%
        \pgfmathprintnumber{\pgfmathresult} dB
};

\end{axis}

\end{tikzpicture}
        \label{fig:rates_bmd_hard}
    }
    \caption{SEs of PAM-$M$ with $M=4,6,8$ and QAM-32 for the peak-power constrained AWGN channel.}
    \label{fig:rates}
\end{figure*}
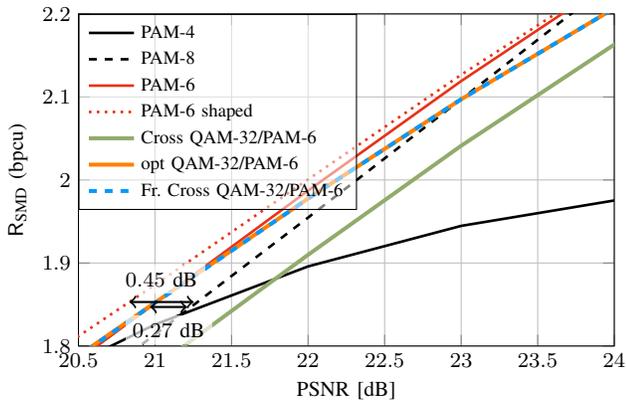
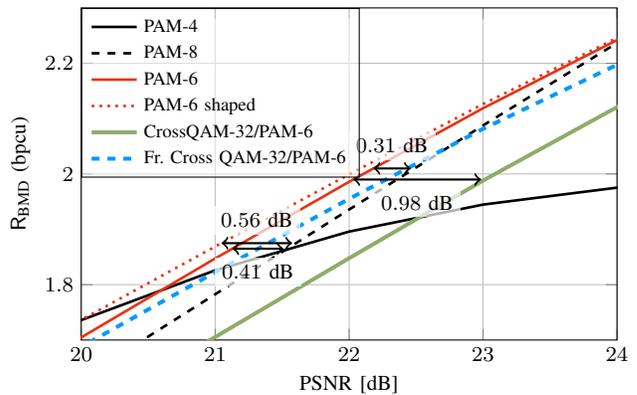
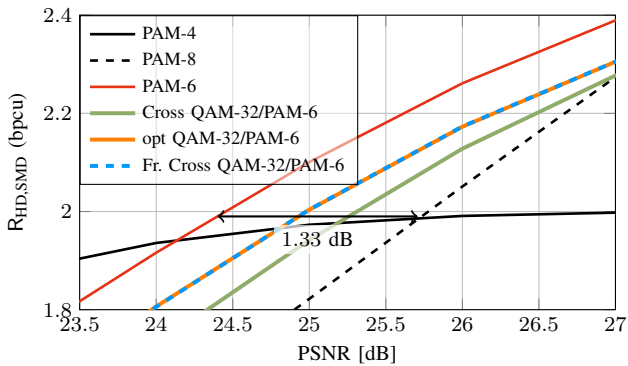
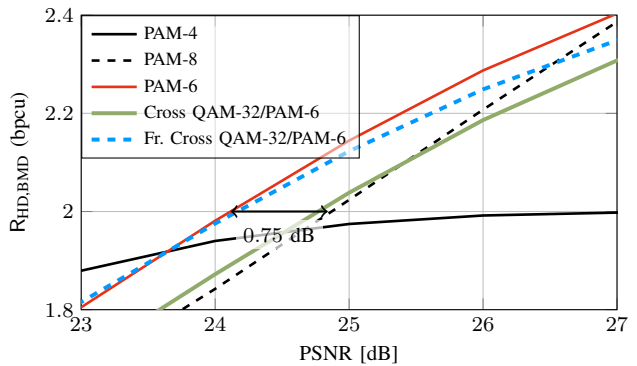

\section{Preliminaries}
\subsection{System Model}
Consider a peak-power constrained additive white Gaussian noise (AWGN) channel with output
\begin{equation}
    Y = X + N
    \label{eq:systemmodel}
\end{equation}
where $X$ is the transmitted signal, and $N$ is AWGN with $N \sim \cN(0, \sigma^2)$. We consider the PAM-$M$ alphabet
\begin{equation}
    \cX = \left\{0, \frac{\sqrt{P_\text{max}}}{M-1}, \dots, (M-1) \frac{\sqrt{P_\text{max}}}{M-1} \right\} \label{eq:cstll}
\end{equation}
where $P_\text{max}$ is the peak transmit power.
The peak-signal-to-noise-power ratio is defined as 
\begin{equation}
    \operatorname{PSNR} = \frac{P_\text{max}}{\sigma^2} .
\end{equation}

\subsection{Spectral Efficiencies}

We study the spectral efficiencies (SEs)
$\Rsmd$, $\Rbmd$, $\Rhdsmd$ and $\Rhdbmd$
under symbol metric decoding (SMD) and bit metric decoding (BMD), and under soft decoding (SD) and hard decoding (HD), respectively. We refer to \cite[Eq. (1)]{bocherer_probabilistic_2019-1} and \cite[Chap.~8]{boecherer2018principles} for more details.  Unless stated otherwise, we use uniformly distributed input symbols with probability $P_X(x) = 1/M,\, \forall\, x\in\mathcal{X}$. 

The SE for SD-SMD is
\begin{align}
    \Rsmd &= \I(X;Y) = \entr(X) - \entr(X|Y) \label{eq:smd_rate} 
\end{align}
where $\I(\cdot; \cdot)$  denotes mutual information, and $\entr(\cdot)$ and $\entr(\cdot|\cdot)$ denote entropy and conditional entropy, respectively.

For BMD, we represent each symbol random variable (RV) $X$ by a sequence of binary RVs $(B_1,\dots,B_m)$, where \mbox{$m=\lceil\log_2M\rceil$}.
The SE for SD-BMD is~\cite{boecherer2015pas}
\begin{align}
    \Rbmd &= \max \left( 0, \,\, \entr(X) - \sum_{k=1}^{m} \entr(B_k|Y) \right). \label{eq:bmd_rate}
\end{align}

For symbol-wise HD, one can achieve
\begin{align}
    \mathsf{R}_\text{HD,SMD} = \max \left( 0, \,\, \entr(X) - (\entr_2(\delta) +\delta \log_2(M-1))
    \right) \label{eq:smd_hd_rate}
\end{align}
where $\entr_2(\cdot)$ is the binary entropy function and ${\delta=\mathrm{Pr}[\hat{X}\!\neq\! X]}$ is the HD symbol error probability with $\hat{X}$ being the symbol-wise hard decisions fed into the detector. 

With BMD the SE under HD is
\begin{equation}
    R_\text{HD,BMD} = \max \left( 0, \,\, \entr(X) - m\entr_2(\varepsilon) \right) \label{eq:bmd_hd_rate}
\end{equation}
where $\varepsilon = 1/m \cdot  \sum_{k=1}^{m} \mathrm{Pr}[\hat B_k \neq B_k]$ is the average hard-decision bit error probability with $\hat{B}_k$ being the bit-wise hard decision of the $k^\text{th}$ bit fed into the detector. 

Fig.~\ref{fig:rates} shows the SEs for SD and HD using SMD and BMD, respectively, for \mbox{PAM-$M$} with $M=4,6,8$ and for \mbox{PAM-$6$} based on two-dimensional \mbox{QAM-$32$} constellations~\cite{ghiasi_2012_investigation,chorchos_2019_pam6}. The \mbox{PAM-$M$} curves are for uniformly-spaced PAM constellations as defined in~\eqref{eq:cstll}. \mbox{PAM-6} using \mbox{32-QAM} constellations is introduced and discussed in Sec.~\ref{sec:QAM-32}. In Sec.~\ref{sec:pampas} we explain how \mbox{PAM-6} can be implemented using PAS.

\begin{figure}
    \centering
    \resizebox{0.9\columnwidth}{!}
    {
    \usetikzlibrary{arrows}

\tikzstyle{point}=[fill,shape=circle,minimum size=4pt,inner sep=0pt]

\begin{tikzpicture}[scale=\columnwidth/11cm,>=latex',yscale=5]
\footnotesize

\node at (6,0) (right2) {$x$};

\node[point] at(0,0.2198) (Point21) {};
\node[point] at(1,0.1297) (Point22) {};
\node[point] at(2,0.1505) (Point23) {};
\node[point] at (3,0.1505) (Point24)  {};
\node[point] at (4,0.1297) (Point25)  {};
\node[point] at(5,0.2198) (Point26) {};

\draw[->] (0,0) -- (right2);
\draw[->] (0,0) -- (0,0.4) node[above,left] {$P_X(x)$};
\draw[-] (0,0) -- (Point21);
\draw[-] (1,0) -- (Point22);
\draw[-] (2,0) -- (Point23);
\draw[-] (3,0) -- (Point24);
\draw[-] (4,0) -- (Point25);
\draw[-] (5,0) -- (Point26);

\draw[dashed] (6,0.2198) -- (-0.1,0.2198) node[left] {0.2198};
\draw[dashed] (6,0.1297) -- (-0.1,0.1297) node[left] {0.1297};
\draw[dashed,TUMBeamerBlue] (-0.1,0.1505) -- (6,0.1505) node[right] {0.1505};

\draw[-] (5,0.01) -- (5,-0.01) node[below] {$\sqrt{P_\mathrm{max}}$};

\end{tikzpicture}
    }
    \vspace{-0.3cm}
    \caption{Optimal distribution $P_X(x)$ of PAM-6 for SD-SMD at a rate of \SI{1.85}{bpcu}.}
    \label{fig:pam6_distribution}
\end{figure}
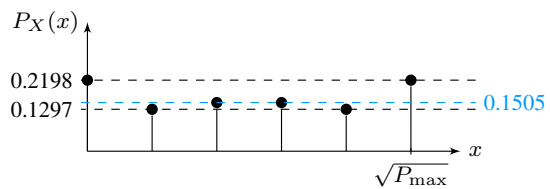

The PAM-$M$ curves in Fig.~\ref{fig:rates} show that each decoding method has a PSNR region where \mbox{PAM-6} outperforms \mbox{PAM-4} and \mbox{PAM-8}. For SD-SMD (Fig.~\ref{fig:rates_smd_soft}), \mbox{PAM-6} gains up to \SI{0.27}{dB} and \SI{0.45}{dB} PSNR compared to \mbox{PAM-8} by using a uniform and optimized input distribution (dotted red), respectively. For a rate of \SI{1.85}{\text{bits/per channel use}} (bpcu), the optimized distribution for SD-SMD is shown in Fig.~\ref{fig:pam6_distribution}. Under SD-BMD (Fig.~\ref{fig:rates_bmd_soft}) \mbox{PAM-6} gains up to \SI{0.41}{dB} and \SI{0.56}{dB} PSNR compared to \mbox{PAM-8} by using a uniform and optimized input distribution, respectively. For HD, the gain is up to \SI{1.33}{dB} and \SI{0.75}{dB} PSNR with SMD (Fig.~\ref{fig:rates_smd_hard}) and BMD (Fig.~\ref{fig:rates_bmd_hard}), respectively.

\section{6-PAM using QAM-32 constellations}\label{sec:QAM-32}
\subsection{Cross QAM-32 constellation}

\begin{figure}[t]
	\centering
	\tikzstyle{point}=[fill,shape=circle,minimum size=3pt,inner sep=0pt]
\begin{tikzpicture}[scale=0.8]
\footnotesize
	\draw[-latex] (0,0) -- (6,0);
	\draw[-latex] (0,0) -- (0,5.5);
	\draw (5,0.2) -- (5,-0.2) node[below] {$\sqrt{P_\text{max}}$};
	\draw (0.2,5) -- (-0.2,5) node[left] {$\sqrt{P_\text{max}}$};
	\node[point,label={[label distance=0.1]-90:{00100}}] at (1,0) {};
	\node[point,label={[label distance=0.1]-90:{00110}}] at (2,0) {};
	\node[point,label={[label distance=0.1]-90:{10110}}] at (3,0) {};
	\node[point,label={[label distance=0.1]-90:{10100}}] at (4,0) {};
	
	\node[point,label={[fill=white,label distance=0.1]-90:{00111}}] at (0,1) {};
	\node[point,label={[label distance=0.1]-90:{00011}}] at (1,1) {};
	\node[point,label={[label distance=0.1]-90:{00010}}] at (2,1) {};
	\node[point,label={[label distance=0.1]-90:{10010}}] at (3,1) {};
	\node[point,label={[label distance=0.1]-90:{10011}}] at (4,1) {};
	\node[point,label={[label distance=0.1]-90:{10111}}] at (5,1) {};
	
	\node[point,label={[fill=white,label distance=0.1]-90:{00101}}] at (0,2) {};
	\node[point,label={[label distance=0.1]-90:{00001}}] at (1,2) {};
	\node[point,label={[label distance=0.1]-90:{00000}}] at (2,2) {};
	\node[point,label={[label distance=0.1]-90:{10000}}] at (3,2) {};
	\node[point,label={[label distance=0.1]-90:{10001}}] at (4,2) {};
	\node[point,label={[label distance=0.1]-90:{10101}}] at (5,2) {};
	
	\node[point,label={[fill=white,label distance=0.1]-90:{01101}}] at (0,3) {};
	\node[point,label={[label distance=0.1]-90:{01001}}] at (1,3) {};
	\node[point,label={[label distance=0.1]-90:{01000}}] at (2,3) {};
	\node[point,label={[label distance=0.1]-90:{11000}}] at (3,3) {};
	\node[point,label={[label distance=0.1]-90:{11001}}] at (4,3) {};
	\node[point,label={[label distance=0.1]-90:{11101}}] at (5,3) {};
	
	\node[point,label={[fill=white,label distance=0.1]-90:{01111}}] at (0,4) {};
	\node[point,label={[label distance=0.1]-90:{01011}}] at (1,4) {};
	\node[point,label={[label distance=0.1]-90:{01010}}] at (2,4) {};
	\node[point,label={[label distance=0.1]-90:{11010}}] at (3,4) {};
	\node[point,label={[label distance=0.1]-90:{11011}}] at (4,4) {};
	\node[point,label={[label distance=0.1]-90:{11111}}] at (5,4) {};
	
	\node[point,label={[label distance=0.1]-90:{01100}}] at (1,5) {};
	\node[point,label={[label distance=0.1]-90:{01110}}] at (2,5) {};
	\node[point,label={[label distance=0.1]-90:{11110}}] at (3,5) {};
	\node[point,label={[label distance=0.1]-90:{11100}}] at (4,5) {};
\end{tikzpicture}
	\vspace{-0.7cm}
	\caption{Cross QAM-32 constellation with quasi-SU labeling~\cite{wesel2001constellation}.}
	\label{fig:qam32_corner}
\end{figure}
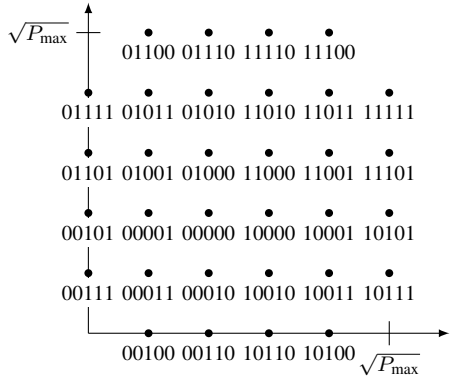

One-dimensional PAM-6 symbols can be created from a two-dimensional \emph{cross~QAM-32} constellation~\cite{ghiasi_2012_investigation, chorchos_2019_pam6} as depicted in Fig.~\ref{fig:qam32_corner}. \mbox{QAM-32} is a simple choice for \mbox{PAM-6} since 5 bits are directly mapped to the \mbox{QAM-32} symbols. We use a quasi symmetric-ultracomposite (SU) labeling structure as introduced in~\cite{wesel2001constellation}. The two-dimensional mapping based on \mbox{QAM-32} is often referred to as \mbox{PAM-6} in the literature, whereas by \mbox{PAM-6} we mean a one-dimensional constellation with 6 uniformly spaced points according to~\eqref{eq:cstll}. Fig.~\ref{fig:rates} shows that cross~\mbox{QAM-32} exhibits a large loss in PSNR compared to PAM-6 under a peak power constraint.  Note that we calculate the rates~\eqref{eq:smd_rate}-\eqref{eq:bmd_hd_rate} with the complex-valued \mbox{32-QAM} constellation and divide the resulting rate by two in order to obtain the rate in bits per \emph{real} channel use.

\subsection{Framed-cross QAM-32 constellation}
Fig.~\ref{fig:pam6_distribution} shows that the optimal input distribution under a peak power constraint has a higher probability at the outer constellation points. 
But this is the opposite as the \mbox{cross~QAM-32} constellation where the inner constellation points are used more often than the outer ones. We thus formulate an optimization problem to obtain the \mbox{QAM-32} constellation that achieves the highest rates. 

\begin{figure}[t]
    \centering
    \tikzstyle{point}=[fill,shape=circle,minimum size=3pt,inner sep=0pt]
\begin{tikzpicture}[scale=1,scale=0.4]
\footnotesize
	\draw[-latex] (0,0) -- (6,0);
	\draw[-latex] (0,0) -- (0,6);
	\draw (5,0.1) -- (5,-0.1) node[below] {$\sqrt{P_\text{max}}$};
	\draw (0.1,5) -- (-0.1,5) node[left] {$\sqrt{P_\text{max}}$};
	\node[point] at (0,0) {};
	\node[point] at (1,0) {};
	\node[point] at (2,0) {};
	\node[point] at (3,0) {};
	\node[point] at (4,0) {};
	\node[point] at (5,0) {};
	
	\node[point] at (0,1) {};
	\node[point] at (1,1) {};
	\node[point] at (2,1) {};
	\node[point] at (4,1) {};
	\node[point] at (5,1) {};
	
	\node[point] at (0,2) {};
	\node[point] at (2,2) {};
	\node[point] at (3,2) {};
	\node[point] at (4,2) {};
	\node[point] at (5,2) {};
	
	\node[point] at (0,3) {};
	\node[point] at (1,3) {};
	\node[point] at (2,3) {};
	\node[point] at (3,3) {};
	\node[point] at (5,3) {};
	
	\node[point] at (0,4) {};
	\node[point] at (1,4) {};
	\node[point] at (3,4) {};
	\node[point] at (4,4) {};
	\node[point] at (5,4) {};
	
	\node[point] at (0,5) {};
	\node[point] at (1,5) {};
	\node[point] at (2,5) {};
	\node[point] at (3,5) {};
	\node[point] at (4,5) {};
	\node[point] at (5,5) {};
\end{tikzpicture}
    \vspace{-0.7cm}
    \caption{Optimal QAM-32 constellation for SMD under a peak power constraint.}
    \label{fig:qam32_grid}
\end{figure}
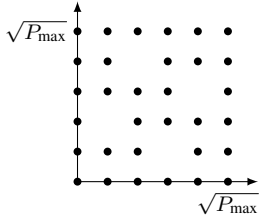

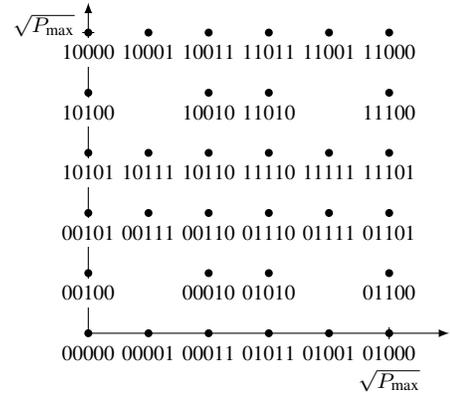
\begin{figure}[t]
	\centering
	\tikzstyle{point}=[fill,shape=circle,minimum size=3pt,inner sep=0pt]
\begin{tikzpicture}[scale=0.8]
\footnotesize
	\draw[-latex] (0,0) -- (6,0);
	\draw[-latex] (0,0) -- (0,5.5);
	\draw (5,0.1) -- (5,-0.1) node[below,yshift=-0.3cm] {$\sqrt{P_\text{max}}$};
	\draw (0.1,5) -- (-0.1,5) node[left,yshift=0.1cm] {$\sqrt{P_\text{max}}$};
	\node[point,label={[fill=white,label distance=0.1]-90:{00000}}] at (0,0) {};
	\node[point,label={[label distance=0.1]-90:{00001}}] at (1,0) {};
	\node[point,label={[label distance=0.1]-90:{00011}}] at (2,0) {};
	\node[point,label={[label distance=0.1]-90:{01011}}] at (3,0) {};
	\node[point,label={[label distance=0.1]-90:{01001}}] at (4,0) {};
	\node[point,label={[label distance=0.1]-90:{01000}}] at (5,0) {};
	
	\node[point,label={[fill=white,label distance=0.1]-90:{00100}}] at (0,1) {};
	\node[point,label={[label distance=0.1]-90:{00010}}] at (2,1) {};
	\node[point,label={[label distance=0.1]-90:{01010}}] at (3,1) {};
	\node[point,label={[label distance=0.1]-90:{01100}}] at (5,1) {};
	
	\node[point,label={[fill=white,label distance=0.1]-90:{00101}}] at (0,2) {};
	\node[point,label={[label distance=0.1]-90:{00111}}] at (1,2) {};
	\node[point,label={[label distance=0.1]-90:{00110}}] at (2,2) {};
	\node[point,label={[label distance=0.1]-90:{01110}}] at (3,2) {};
	\node[point,label={[label distance=0.1]-90:{01111}}] at (4,2) {};
	\node[point,label={[label distance=0.1]-90:{01101}}] at (5,2) {};
	
	\node[point,label={[fill=white,label distance=0.1]-90:{10101}}] at (0,3) {};
	\node[point,label={[label distance=0.1]-90:{10111}}] at (1,3) {};
	\node[point,label={[label distance=0.1]-90:{10110}}] at (2,3) {};
	\node[point,label={[label distance=0.1]-90:{11110}}] at (3,3) {};
	\node[point,label={[label distance=0.1]-90:{11111}}] at (4,3) {};
	\node[point,label={[label distance=0.1]-90:{11101}}] at (5,3) {};
	
	\node[point,label={[fill=white,label distance=0.1]-90:{10100}}] at (0,4) {};
	\node[point,label={[label distance=0.1]-90:{10010}}] at (2,4) {};
	\node[point,label={[label distance=0.1]-90:{11010}}] at (3,4) {};
	\node[point,label={[label distance=0.1]-90:{11100}}] at (5,4) {};
	
	\node[point,label={[fill=white,label distance=0.1]-90:{10000}}] at (0,5) {};
	\node[point,label={[label distance=0.1]-90:{10001}}] at (1,5) {};
	\node[point,label={[label distance=0.1]-90:{10011}}] at (2,5) {};
	\node[point,label={[label distance=0.1]-90:{11011}}] at (3,5) {};
	\node[point,label={[label distance=0.1]-90:{11001}}] at (4,5) {};
	\node[point,label={[label distance=0.1]-90:{11000}}] at (5,5) {};
\end{tikzpicture}
	\vspace{-0.7cm}
	\caption{Framed-cross QAM-32 constellation with Gray labeling.}
	\label{fig:qam32_framed_cross}
\end{figure}

Define the \mbox{QAM-36} constellation
\begin{align}
    \mathcal{X}_{36} = \left\{ a + \mathrm{j} b \mid a,b \in \mathcal{X}  \right\}
\end{align}
where we use $\mathcal{X}$ from~\eqref{eq:cstll} with $M=6$. Now maximize the SMD-SD rate over all 32-symbol subsets $\tilde{\mathcal{X}}$ of $\mathcal{X}_{36}$: 
\begin{align}
    \mathcal{X}_{32,\mathrm{SMD}}^\star = \argmax_{\substack{\tilde{\mathcal{X}} \subset \mathcal{X}_{36} \\ |\tilde{\mathcal{X}}| = 32}} \Rsmd
    .
    \label{eq:opt_smd_32_symbol_subset}
\end{align}
The optimal constellation $\mathcal{X}_{32,\mathrm{SMD}}^\star $ according to~\eqref{eq:opt_smd_32_symbol_subset} is depicted in Fig.~\ref{fig:qam32_grid}.
This constellation does not allow Gray labeling and finding the best bit labeling under BMD seems difficult. Instead, we modify~\eqref{eq:opt_smd_32_symbol_subset} by taking the BMD rate from~\eqref{eq:bmd_rate} as the new objective function. Furthermore, we impose a Gray labeling constraint were the first two bits of the label correspond to the quadrant of the constellation. The result is shown in Fig.~\ref{fig:qam32_framed_cross} and we denote this constellation as \emph{framed-cross} \mbox{QAM-32}.

The information rates of the optimized \mbox{QAM-32} and framed-cross \mbox{QAM-32} constellations are shown in Fig.~\ref{fig:rates}. Fig.~\ref{fig:rates_smd_soft} and~\ref{fig:rates_smd_hard} show that there is no visible PSNR loss between the suboptimal framed-cross constellation and optimized \mbox{QAM-32}. Furthermore, framed-cross QAM-32 always outperforms cross \mbox{QAM-32}. For SD-SMD and HD-BMD, framed-cross \mbox{QAM-32} achieves  similar rates as \mbox{PAM-6} if the rates are low. For HD-SMD and SD-BMD, the rate of framed-cross \mbox{QAM-32} lies between that of \mbox{PAM-6} and cross \mbox{QAM-32}. 

\section{$M$-PAM using PAS}\label{sec:pampas}
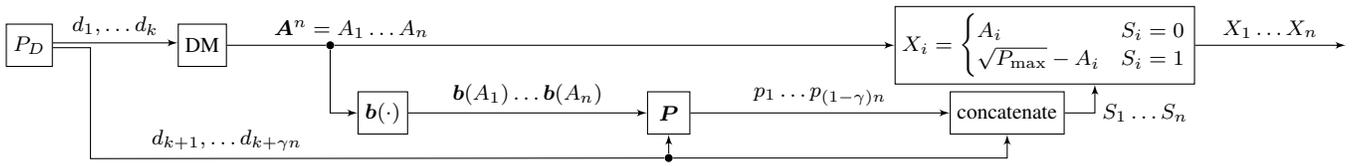
\begin{figure*}
    \centering
    \tikzstyle{block} = [draw,minimum size=2em]
	\tikzstyle{prod} = [draw,circle,inner sep=0.1mm,minimum size=0.5em]
	\tikzstyle{point}=[fill,shape=circle,minimum size=3pt,inner sep=0pt]
	
	\begin{tikzpicture}[>=latex',yscale=0.6]
	\footnotesize
	\node[block] at (-0.2,0) (dm) {DM};
	\node[block] at (-2.5,0) (source) {$P_D$};
	\node[point] at (1.5,0) (point1) {};
	\node[point] at(6,-2.5) (point2) {};
	\node[block] at (2.2,-1.5) (labeling) {$\boldsymbol{b} (\cdot)$};
	\node[block] at (6,-1.5) (encoder) {$\boldsymbol{P}$};
	\node[block,align=center] at (10.5,-1.5) (unlabeling) {concatenate};
	\node[block,align=center] at (11,0) (product1) {$X_i = \begin{cases}A_i & S_i=0 \\ \sqrt{P_\mathrm{max}}-A_i & S_i=1\end{cases}$};

	\draw (dm) -- (point1) node[midway,above,xshift=1cm] {$\boldsymbol{A}^{n} = A_1\ldots A_{n}$};
	\draw[->] (point1) |- (labeling);
	\draw[->] (labeling) -- (encoder) node[midway,above] {$\boldsymbol{b}(A_1)\ldots \boldsymbol{b}(A_{n})$};
	\draw[->] (encoder) -- (unlabeling) node[midway,above] {$p_1\ldots p_{(1-\gamma)n}$};
	\draw[->] (unlabeling) -| ($(product1.south west)!2/3!(product1.south east)$) node[midway,right] {$S_1\ldots S_{n}$};
	\draw[->] (point1) -- (product1);
	\draw[->] (product1) -- (15,0) node[midway,above] {$X_1\ldots X_n$};

	\draw[->] (point2) -- (10.5,-2.5) -- (unlabeling);
	
	\draw[->] ($(source.north east)!4/9!(source.south east)$) -- ($(dm.north west)!4/9!(dm.south west)$) node[midway,above] {$d_1,\ldots d_{k}$};
	\draw ($(source.north east)!5/9!(source.south east)$) --++ (0.5,0) --++ (0,-2.46) -- (point2) node[midway,right,above,xshift=-2cm] {$d_{k+1},\ldots d_{k+\gamma n}$};
	\draw[->] (point2) -- (encoder);

	\end{tikzpicture}
    \vspace{-0.5cm}
    \caption{Block diagram for probabilistic amplitude shaping (PAS).}
    \label{fig:pas}
\end{figure*}
PAS was designed for distributions $P_X$ that are symmetric around zero~\cite{boecherer2015pas}. 
However, PAS can be extended to any distribution with a symmetry.
We propose a PAS scheme for PAM-$M$ distributions for even integers $M>2$. The number of bits needed to label the constellation points is $m = \lceil \log_2M \rceil$. The same idea is used in~\cite{buchali2018flexible} to construct flexible QAM constellations. 

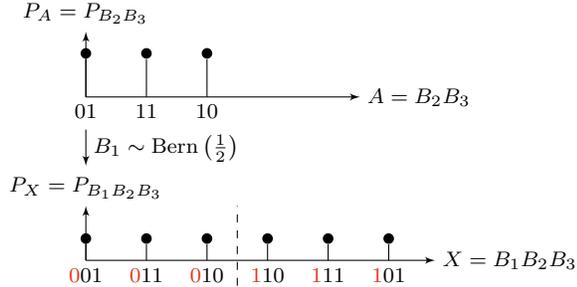
\begin{figure}
    \centering
    \usetikzlibrary{arrows}

\tikzstyle{point}=[fill,shape=circle,minimum size=4pt,inner sep=0pt]

\begin{tikzpicture}[scale=\columnwidth/11cm,>=latex',yscale=0.6]
\footnotesize
\node at (5.5,0) (right) {$A = B_2B_3$};
\node at (0,2.3) (oben) {$P_A = P_{B_2B_3}$};
\node[point] at (0,1.2) (Point1)  {};
\node[point] at(1,1.2) (Point2) {};
\node[point] at(2,1.2) (Point3) {};

\def \labelShift{-0.4}
\draw[->] (0,0) -- (right);
\draw[->] (0,0) -- (oben);
\draw[-] (0,0) -- (Point1);
\draw[-] (1,0) -- (Point2);
\draw[-] (2,0) -- (Point3);

\node at(0,\labelShift) {$01$};
\node at(1,\labelShift) {$11$};
\node at(2,\labelShift) {$10$};

\def\shift {-4.5};
\node[point] at(0,\shift + 0.6) (Point21) {};
\node[point] at(1,\shift + 0.6) (Point22) {};
\node[point] at(2,\shift + 0.6) (Point23) {};
\node[point] at (3,\shift+ 0.6) (Point24)  {};
\node[point] at (4,\shift+ 0.6) (Point25)  {};
\node[point] at(5,\shift + 0.6) (Point26) {};

\node at (7,\shift) (right2) {$X = B_1B_2B_3$};
\node at (0,\shift + 2) (oben2) {$P_X = P_{B_1B_2B_3}$};

\draw[->] (0,\shift) -- (right2);
\draw[->] (0,\shift) -- (oben2);
\draw[-] (0,\shift) -- (Point21);
\draw[-] (1,\shift) -- (Point22);
\draw[-] (2,\shift) -- (Point23);
\draw[-] (3,\shift) -- (Point24);
\draw[-] (4,\shift) -- (Point25);
\draw[-] (5,\shift) -- (Point26);

\node at(0,\labelShift+\shift) {${\color{TUMBeamerRed}0}01$};
\node at(1,\labelShift+\shift) {${\color{TUMBeamerRed}0}11$};
\node at(2,\labelShift+\shift) {${\color{TUMBeamerRed}0}10$};
\node at(3,\labelShift+\shift) {${\color{TUMBeamerRed}1}10$};
\node at(4,\labelShift+\shift) {${\color{TUMBeamerRed}1}11$};
\node at(5,\labelShift+\shift) {${\color{TUMBeamerRed}1}01$};

\draw[->] (0,-0.9) -- (0,-1.9) node[midway,right] {$B_1 \sim \mathrm{Bern}\left(\frac{1}{2}\right)$ };

\draw[dashed] (2.5,\shift-0.7) -- (2.5,\shift+1.5);

\end{tikzpicture}
    \vspace{-0.7cm}
    \caption{Generating PAM-$6$ symbols using PAS: PAM-$3$ symbols are generated using a DM and labelled with two bits. Uniform parity bits are used to extend the PAM-$3$ symbols to PAM-$6$ symbols with a three bit label.}
    \label{fig:pas_example}
\end{figure}

The PAS scheme is illustrated in Fig.~\ref{fig:pas}. A source $P_D$ outputs uniformly distributed data bits \mbox{$d_i$, $i=1,2,\dots,k+\gamma n$}. A subsequence $d_1,\ldots,d_k$ is matched to a sequence $a_1,a_2,\ldots,a_n$ with a desired distribution $P_A(.)$ by a distribution matcher (DM). We use constant composition distribution matching (CCDM)~\cite{schulte2015constant} in this paper, but other DMs can be used as well.  The ``amplitude'' constellation $\mathcal{A}$ is the set of the first $M/2$ symbols of the PAM-$M$ constellation $\mathcal{X}$:
\begin{align}
\mathcal{A} = \left\{ i\cdot \frac{\sqrt{P_\mathrm{max}}}{M-1}  \right\}_{i=0}^{M/2-1}.
\end{align}
Every point $a\in\mathcal{A}$ has a binary label $\boldsymbol{b}(a) = (b_{2},\ldots,b_{m})$ of length~$m-1$. We use subsets of binary reflected Gray codes (BRGC)~\cite{frank1953pulse} of length~$m-1$, where we discard the first ${2^{m-1}-M/2}$ codewords from the usual Gray code (compare the labeling of $A=B_2B_3$ in Fig.~\ref{fig:pas_example}).

The FEC code $\mathcal{C}$ is binary, and has length $m\cdot n$ and dimension ${(m-1+\gamma)n}$ with $0\leq\gamma\leq 1$. Let $\boldsymbol{P}$ be the parity-generating matrix of a systematic generator matrix $\boldsymbol{G} = [\boldsymbol{I}_k|\boldsymbol{P}]$. Systematic encoding multiplies the bit string of length ${(m-1+\gamma)n}$ with $\boldsymbol{P}$, where the bit string consists of the binary labels and $\gamma n$ additional uniformly distributed bits $d_{k+1},\ldots,d_{k+\gamma n}$ from the source. This results in $(1-\gamma)n$ parity bits $p_1,\ldots,p_{(1-\gamma)n}$ that we assume to be uniformly distributed at the decoder~\cite{boecherer_2014_labeling}. Together with the $\gamma n$ bits $d_{k+1},\ldots,d_{k+\gamma n}$, we have $n$ bits that map the $M/2$ symbols of $\mathcal{A}$ to either $\mathcal{A}$ or $\mathcal{X}\setminus\mathcal{A}$.

The SE of this scheme is given by \eqref{eq:smd_rate}-\eqref{eq:bmd_hd_rate}. The overall SE including DM and FEC is
\begin{align}
    \mathrm{SE} = R_{\mathrm{DM}}+1 - m(1-R_{\mathrm{FEC}})
\end{align}
where $R_{\mathrm{DM}}$ denotes the rate of the DM and $R_{\mathrm{FEC}}$ the code rate of the FEC code. The relation to  $\entr(X)$ in \eqref{eq:smd_rate}-\eqref{eq:bmd_hd_rate} is
\begin{align}
    R_{\mathrm{DM}} + 1 \leq \entr(A)+1 = \entr(X)
\end{align}
where $\entr(A) \leq \log_2(M)-1$, with equality iff $A$ is uniformly distributed. The FEC overhead $m(1-R_{\mathrm{FEC}})$ is larger or equal to the terms that are substracted from $\entr(X)$ in \eqref{eq:smd_rate}-\eqref{eq:bmd_hd_rate}. The difference between $R_{\mathrm{DM}}$ and $\entr(A)$ is called the rate loss, and it approaches zero asymptotically in the DM output length~\cite{schulte2015constant}.

\section{Comparison / Numerical Results}
We evaluate PAM-6 with PAS and compare it to PAM-6 generated by 32-QAM, and to PAM-8. We further compare SEs for both HD and SD and provide coded results employing low-density parity-check (LDPC) codes. We perform 200 and 20 belief-propagation iterations with a full sum-product update rule for SD and HD, respectively.

\subsection{SD-BMD using 5G LDPC codes}
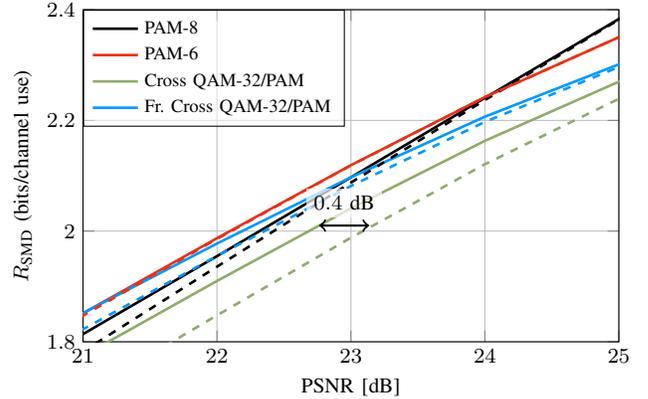
\begin{figure}
	\centering
	\begin{tikzpicture} 
\footnotesize
\begin{axis}[
xlabel={PSNR [\si{dB}]},
width=.48\textwidth, 
height=6cm,
ylabel={$R_\tsmd$ (bits/channel use)},
grid=both,
legend cell align={left},
 legend style ={at={(0,1.0)},anchor=north west, fill opacity=0.5, text opacity=1},
ymin=1.8,
ymax=2.4,
xmin=21,
xmax=25,
]

\addplot[name path global=pam8smd,line width=1,black] table[x=psnr,y=rate] {results/rates_8pam_smd_uniform.txt};
\addlegendentry{\scriptsize PAM-8};

\addplot[name path global=pam6smd,line width=1,TUMBeamerRed] table[x=psnr,y=rate] {results/rates_6pam_smd_uniform.txt};
\addlegendentry{\scriptsize PAM-6};

\addplot[name path global=qam32smd,line width=1,TUMBeamerGreen] table[x=psnr,y=rate] {results/rates_cross_32qam_smd_uniform.txt};
\addlegendentry{\scriptsize Cross QAM-32/PAM};

\addplot[name path global=qam32smd_framed_cross,line width=1,TUMBeamerBlue] table[x=psnr,y=rate] {results/rates_framed_cross_32qam_smd_uniform.txt};
\addlegendentry{\scriptsize Fr. Cross QAM-32/PAM};

\addplot[name path global=pam8bmd,line width=1,black,dashed] table[x=psnr,y=rate] {results/rates_8pam_bmd_uniform.txt};

\addplot[name path global=pam6bmd,line width=1,TUMBeamerRed,dashed] table[x=psnr,y=rate] {results/rates_6pam_bmd_uniform.txt};

\addplot[name path global=qam32bmd,line width=1,TUMBeamerGreen,dashed] table[x=psnr,y=rate] {results/rates_cross_32qam_bmd_uniform.txt};

\addplot[name path global=qam32bmd_framed_cross,line width=1,TUMBeamerBlue,dashed] table[x=psnr,y=rate] {results/rates_framed_cross_32qam_bmd_uniform.txt};

\path[name path global=line] (axis cs:\pgfkeysvalueof{/pgfplots/xmin},2.01) -- (axis cs: \pgfkeysvalueof{/pgfplots/xmax},2.01);
\path[name intersections={of=line and qam32smd, name=p1}, name intersections={of=line and qam32bmd, name=p2}];
\draw[<->,thick] let \p1=(p1-1), \p2=(p2-1) in (p1-1) -- ([xshift=-0.05cm]p2-1) node [above,midway,yshift=0.1cm,fill=white, fill opacity=0.7,text opacity=1, draw opacity=1] {%
	\pgfplotsconvertunittocoordinate{x}{\x1}%
	\pgfplotscoordmath{x}{datascaletrafo inverse to fixed}{\pgfmathresult}%
	\edef\valueA{\pgfmathresult}%
	\pgfplotsconvertunittocoordinate{x}{\x2}%
	\pgfplotscoordmath{x}{datascaletrafo inverse to fixed}{\pgfmathresult}%
	\pgfmathparse{\pgfmathresult - \valueA}%
	\pgfmathprintnumber{\pgfmathresult} dB
};

\end{axis}

\end{tikzpicture}
	\vspace{-0.7cm}
	\caption{SEs under BMD (dashed) and SMD (solid) soft-decoding.}
	\label{fig:rates_soft}
\end{figure}

Compared to SMD, BMD with soft-decoding exhibits a loss in PSNR, as BMD uses a mismatched decoding metric instead of $P_{X|Y}$. The PSNR loss for \mbox{PAM-6}, \mbox{PAM-8} and \mbox{QAM-32/PAM} is shown in Fig.~\ref{fig:rates_soft} as the gap between the dashed (BMD) and solid (SMD) curves. The \mbox{PAM-6} loss is negligible in the PSNR region of interest, whereas a $\SI{0.4}{dB}$ loss between SMD and BMS is evident for cross \mbox{QAM-32}. 

In the following, we compare the schemes at a rate of \SI{2}{bpcu}. As shown in Fig.~\ref{fig:rates_bmd_soft}, \mbox{PAM-6} gains \SI{0.31}{dB} over \mbox{PAM-8} and \SI{0.98}{dB} over cross \mbox{QAM-32}. \mbox{PAM-6} outperforms \mbox{PAM-8} up to a SE of $R=\SI{2.25}{\text{bpcu}}$. The cross \mbox{QAM-32} SE is lower than \mbox{PAM-6} and \mbox{PAM-8} over the entire PSNR range.

The benefits of PAS-PAM-6 and framed-cross QAM-32 over cross QAM-32 are twofold: 
\begin{enumerate}
    \item The uniform distribution of PAS-PAM-6 and the non-uniform distribution implied by framed-cross QAM-32 are closer to the optimal distribution (cf. Fig.~\ref{fig:pam6_distribution}) under a peak-power constraint than the distribution implied by cross QAM-32.
    \item The BMD loss is minimized due to the ``true'' Gray labeling of PAS-PAM-6 and framed-cross QAM-32. 
\end{enumerate}
PAS-PAM-6 has an additional small gain: the entropy of \mbox{PAM-6} is $\log_2(6) \approx \SI{2.58}{\text{bpcu}}$ compared to $\SI{2.5}{\text{bpcu}}$ for the \mbox{QAM-32} constellations.

\begin{figure}
	\centering
	\begin{tikzpicture} 
\footnotesize
\begin{axis}[
xlabel={PSNR [\si{dB}]},
width=.48\textwidth, 
height=5.5cm,
ylabel={FER},
ymode=log,
grid=both,
legend cell align={left},
 legend style ={at={(1,1.0)},anchor=north east, fill opacity=0.5, text opacity=1},
xmin=22.5,
ymax=1,
ymin=1e-5,
]

\addplot[name path global=pam8,line width=1,black] table[x=psnr,y=fer] {results/simulation_results/fer_LDPC_8PAM_n=3000_R=2.00_it=200.txt};
\addlegendentry{\scriptsize PAM-8};

\addplot[name path global=pam6,line width=1,TUMBeamerRed] table[x=psnr,y=fer] {results/simulation_results/fer_LDPC_6PAM_PAS_n=3000_R=2.00_it=200.txt};
\addlegendentry{\scriptsize PAS PAM-6};

\addplot[name path global=pam6shaped,line width=1,TUMBeamerRed,dotted] table[x=psnr,y=fer] {results/simulation_results/fer_new_LDPC_6PAM_PAS_shaped_n=3000_R=2.00_it=200.txt};
\addlegendentry{\scriptsize PAS PAM-6, optimized $P_X$};

\addplot[name path global=qam32,line width=1,TUMBeamerGreen] table[x=psnr,y=fer] {results/simulation_results/fer_LDPC_32QAM_corner_n=3000_R=2.00_it=200.txt};
\addlegendentry{\scriptsize Cross QAM-32/PAM-6};

\addplot[name path global=qam32framedcross,line width=1,TUMBeamerBlue] table[x=psnr,y=fer] {results/simulation_results/fer_LDPC_32QAM_cross_n=3000_R=2.00_it=200.txt};
\addlegendentry{\scriptsize Fr. Cross QAM-32/PAM-6};

\path[name path global=line] (axis cs:\pgfkeysvalueof{/pgfplots/xmin},1.2e-3) -- (axis cs: \pgfkeysvalueof{/pgfplots/xmax},1.2e-3);
\path[name intersections={of=line and pam6, name=p1}, name intersections={of=line and pam8, name=p2}];
\draw[<->,thick] let \p1=(p1-1), \p2=(p2-1) in (p1-1) -- ([xshift=-0.05cm]p2-1) node [above,midway,yshift=0.1cm,fill=white, fill opacity=0.7,text opacity=1, draw opacity=1] {%
	\pgfplotsconvertunittocoordinate{x}{\x1}%
	\pgfplotscoordmath{x}{datascaletrafo inverse to fixed}{\pgfmathresult}%
	\edef\valueA{\pgfmathresult}%
	\pgfplotsconvertunittocoordinate{x}{\x2}%
	\pgfplotscoordmath{x}{datascaletrafo inverse to fixed}{\pgfmathresult}%
	\pgfmathparse{\pgfmathresult - \valueA}%
	\pgfmathprintnumber{\pgfmathresult} dB
};

\path[name path global=line] (axis cs:\pgfkeysvalueof{/pgfplots/xmin},7e-4) -- (axis cs: \pgfkeysvalueof{/pgfplots/xmax},7e-4);
\path[name intersections={of=line and pam6, name=p1}, name intersections={of=line and qam32framedcross, name=p2}];
\draw[<->,thick] let \p1=(p1-1), \p2=(p2-1) in (p1-1) -- ([xshift=-0.05cm]p2-1) node [below,midway,yshift=-0.1cm,fill=white, fill opacity=0.7,text opacity=1, draw opacity=1] {%
	\pgfplotsconvertunittocoordinate{x}{\x1}%
	\pgfplotscoordmath{x}{datascaletrafo inverse to fixed}{\pgfmathresult}%
	\edef\valueA{\pgfmathresult}%
	\pgfplotsconvertunittocoordinate{x}{\x2}%
	\pgfplotscoordmath{x}{datascaletrafo inverse to fixed}{\pgfmathresult}%
	\pgfmathparse{\pgfmathresult - \valueA}%
	\pgfmathprintnumber{\pgfmathresult} dB
};

\path[name path global=line] (axis cs:\pgfkeysvalueof{/pgfplots/xmin},9e-4) -- (axis cs: \pgfkeysvalueof{/pgfplots/xmax},9e-4);
\path[name intersections={of=line and pam6, name=p1}, name intersections={of=line and qam32, name=p2}];
\draw[<->,thick] let \p1=(p1-1), \p2=(p2-1) in (p1-1) -- ([xshift=-0.05cm]p2-1) node [below,midway,yshift=-0.1cm,fill=white, fill opacity=0.7,text opacity=1, draw opacity=1] {%
	\pgfplotsconvertunittocoordinate{x}{\x1}%
	\pgfplotscoordmath{x}{datascaletrafo inverse to fixed}{\pgfmathresult}%
	\edef\valueA{\pgfmathresult}%
	\pgfplotsconvertunittocoordinate{x}{\x2}%
	\pgfplotscoordmath{x}{datascaletrafo inverse to fixed}{\pgfmathresult}%
	\pgfmathparse{\pgfmathresult - \valueA}%
	\pgfmathprintnumber{\pgfmathresult} dB
};

\end{axis}

\end{tikzpicture}
	\vspace{-0.7cm}
	\caption{Performance comparison of PAM-6 schemes at a SE of $\SI{2}{bpcu}$ under SD-BMD with $n=3000$ channel uses per frame.}
	\label{fig:soft_results}
\end{figure}
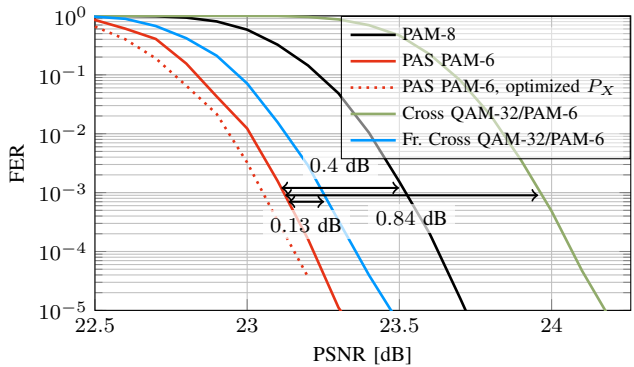

The SEs give reasonable predictions of the PSNR gains with implemented codes, as shown by the frame error rate (FER) curves in Fig.~\ref{fig:soft_results}. We use LDPC codes as described in the 5G new radio standard \cite{3gpp-ts-38.212-v15.0.0-17-12a} with a block length of $n=3000$ channel uses (3000~ \mbox{PAM-6/8} symbols and 1500~\mbox{QAM-32} symbols, respectively). The overall SE is $\SI{2}{\text{bpcu}}$. The code parameters are summarized in Table~\ref{tab:simparams}. The rate loss of the CCDM is \SI{0.004}{\text{bpcu}}. At a FER of $10^{-3}$, PAM-$6$ using PAS gains $\SI{0.4}{dB}$ and $\SI{0.84}{dB}$  compared to PAM-$8$ and cross QAM-$32$, respectively. The FER of framed-cross QAM-32 is within \SI{0.13}{dB} of PAM-6. These results are in line with the SEs from Fig.~\ref{fig:rates_bmd_soft}.

\begin{table}
\centering
 \begin{tabular}{||r | l| l| l||} 
 \hline
 Scenario & code length & coderate $R_{\mathrm{FEC}}$ & $\gamma n$ \\ [0.5ex] 
 \hline\hline
 PAS PAM-6 & \SI{9000}{bits} & $7257/9000\approx 0.8$ & \SI{1257}{bits}\\ 
 PAM-8 & \SI{9000}{bits} & $2/3$ & \\
 QAM-32/PAM-6 & \SI{7500}{bits} & $0.8$ &  \\
 
 \hline
 \end{tabular}
 \vspace{0.2cm}
 \caption{Code parameters.}
 \vspace{-0.5cm}
 \label{tab:simparams}
\end{table}

\subsection{HD using 5G LDPC Codes}
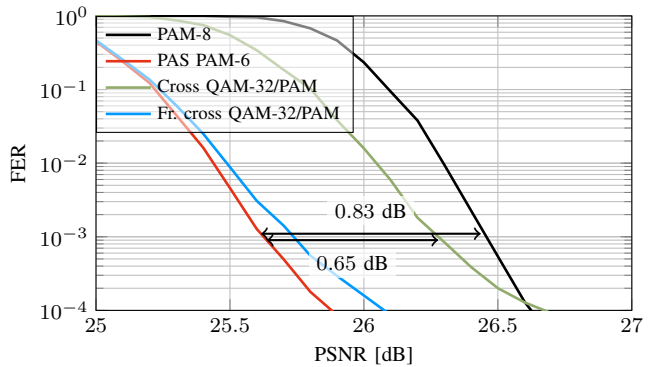
\begin{figure}[t]
	\centering
	\begin{tikzpicture} 
\footnotesize
\begin{axis}[
xlabel={PSNR [\si{dB}]},
width=.48\textwidth, 
height=5.5cm,
ylabel={FER},
ymode=log,
grid=both,
legend cell align={left},
 legend style ={at={(0,1.0)},anchor=north west, fill opacity=0.5, text opacity=1},
xmin=25,
ymax=1,
ymin=1e-4,
xmax=27
]

\addplot[name path global=pam8,line width=1,black] table[x=psnr,y=fer] {results/simulation_results/hd_fer_LDPC_8PAM_n=3000_R=2.00_it=20.txt};
\addlegendentry{\scriptsize PAM-8};

\addplot[name path global=pam6,line width=1,TUMBeamerRed] table[x=psnr,y=fer] {results/simulation_results/hd_fer_LDPC_6PAM_PAS_n=3000_R=2.00_it=20.txt};
\addlegendentry{\scriptsize PAS PAM-6};

\addplot[name path global=qam32,line width=1,TUMBeamerGreen] table[x=psnr,y=fer] {results/simulation_results/hd_fer_LDPC_32QAM_corner_n=3000_R=2.00_it=20.txt};
\addlegendentry{\scriptsize Cross QAM-32/PAM};

\addplot[name path global=qam32framed_cross,line width=1,TUMBeamerBlue] table[x=psnr,y=fer] {results/simulation_results/hd_fer_LDPC_32QAM_cross_n=3000_R=2.00_it=20.txt};
\addlegendentry{\scriptsize Fr. cross QAM-32/PAM};

\path[name path global=line] (axis cs:\pgfkeysvalueof{/pgfplots/xmin},1.1e-3) -- (axis cs: \pgfkeysvalueof{/pgfplots/xmax},1.1e-3);
\path[name intersections={of=line and pam6, name=p1}, name intersections={of=line and pam8, name=p2}];
\draw[<->,thick] let \p1=(p1-1), \p2=(p2-1) in (p1-1) -- ([xshift=-0.05cm]p2-1) node [above,midway,yshift=0.1cm,fill=white, fill opacity=0.7,text opacity=1, draw opacity=1] {%
	\pgfplotsconvertunittocoordinate{x}{\x1}%
	\pgfplotscoordmath{x}{datascaletrafo inverse to fixed}{\pgfmathresult}%
	\edef\valueA{\pgfmathresult}%
	\pgfplotsconvertunittocoordinate{x}{\x2}%
	\pgfplotscoordmath{x}{datascaletrafo inverse to fixed}{\pgfmathresult}%
	\pgfmathparse{\pgfmathresult - \valueA}%
	\pgfmathprintnumber{\pgfmathresult} dB
};

\path[name path global=line] (axis cs:\pgfkeysvalueof{/pgfplots/xmin},9e-4) -- (axis cs: \pgfkeysvalueof{/pgfplots/xmax},9e-4);
\path[name intersections={of=line and pam6, name=p1}, name intersections={of=line and qam32, name=p2}];
\draw[<->,thick] let \p1=(p1-1), \p2=(p2-1) in (p1-1) -- ([xshift=-0.05cm]p2-1) node [below,midway,yshift=-0.1cm,fill=white, fill opacity=0.7,text opacity=1, draw opacity=1] {%
	\pgfplotsconvertunittocoordinate{x}{\x1}%
	\pgfplotscoordmath{x}{datascaletrafo inverse to fixed}{\pgfmathresult}%
	\edef\valueA{\pgfmathresult}%
	\pgfplotsconvertunittocoordinate{x}{\x2}%
	\pgfplotscoordmath{x}{datascaletrafo inverse to fixed}{\pgfmathresult}%
	\pgfmathparse{\pgfmathresult - \valueA}%
	\pgfmathprintnumber{\pgfmathresult} dB
};

\end{axis}

\end{tikzpicture}
	\vspace{-0.7cm}
	\caption{Performance of PAS-PAM-$6$ at a SE of $\SI{2}{\text{bpcu}}$ under HD-BMD using 5G LDPC codes with $n=3000$ channel uses per frame.}
	\label{fig:hd_results}
\end{figure}

SD FEC codes may be too complex for high-throughput applications such as IM/DD for short-reach fiber-optic links. Instead, one often uses HD. SEs for HD are depicted in Fig.~\ref{fig:rates}. Observe that BMD can improve on SMD under HD, see Figs.~\ref{fig:rates_smd_hard} and \ref{fig:rates_bmd_hard}, see~\cite{sheikh2017achievable}. The reason is that BMD works with the binary Hamming distance that preserves part of the Euclidean distance, while the $M$-ary Hamming metric for SMD does not. To validate the results, Fig.~\ref{fig:hd_results} shows the performance of 5G LDPC codes with HD. We make bit-wise hard decisions on the channel outputs and assign the values $+a$ and $-a$ to the LLRs if the decision is 0 or 1, respectively.
The value $a$ is the same for all bit-levels and symbols and the optimal value of $a$ is found with a line search. We use the same simulation parameters as for SD decoding, i.e., $n=3000$ channel uses and a target SE of $\SI{2}{\text{bpcu}}$. PAS-PAM-$6$ gains $\SI{0.83}{dB}$ and $\SI{0.65}{dB}$ over \mbox{PAM-$8$} and \mbox{QAM-$32$} based \mbox{PAM-$6$}, respectively. This confirms the SE results from Fig.~\ref{fig:rates_bmd_hard}. 

We remark that there are codes that perform better under HD-BMD than 5G LDPC codes. The aim here is to show that information rate gains predict the real coding gains, and 5G LDPC codes allow a good rate adaption to approach the rates described in Tab.~\ref{tab:simparams}.

\section{Conclusions}
PAM-6 exhibits PSNR gains compared to \mbox{PAM-4} and \mbox{PAM-8} when transmitting over an AWGN channel with a peak power constraint. We proposed a PAS scheme to generate \mbox{PAM-6} symbols and show that this schemes outperforms \mbox{PAM-8} and a conventional \mbox{PAM-6} scheme that is based on a \mbox{2-D} to \mbox{1-D} mapping of a complex-valued cross-shaped \mbox{QAM-32} constellation.
Additionally, we introduced an optimized \mbox{QAM-32} constellation with the shape of a framed cross that achieves similar PSNR gains as the previously mentioned PAS scheme. The binary code length of framed-cross \mbox{QAM-32} is less than for the PAS scheme, which might be beneficial in cheap hardware. The additional gains of the PAS scheme do come at the cost of increasing computational complexity due to the DM. 
Using the DM to optimize the input distribution achieves further PSNR gains.

\bibliographystyle{IEEEtran}
\bibliography{IEEEabrv,references}

\begin{thebibliography}{10}
\providecommand{\url}[1]{#1}
\csname url@samestyle\endcsname
\providecommand{\newblock}{\relax}
\providecommand{\bibinfo}[2]{#2}
\providecommand{\BIBentrySTDinterwordspacing}{\spaceskip=0pt\relax}
\providecommand{\BIBentryALTinterwordstretchfactor}{4}
\providecommand{\BIBentryALTinterwordspacing}{\spaceskip=\fontdimen2\font plus
\BIBentryALTinterwordstretchfactor\fontdimen3\font minus
  \fontdimen4\font\relax}
\providecommand{\BIBforeignlanguage}[2]{{%
\expandafter\ifx\csname l@#1\endcsname\relax
\typeout{** WARNING: IEEEtran.bst: No hyphenation pattern has been}%
\typeout{** loaded for the language `#1'. Using the pattern for}%
\typeout{** the default language instead.}%
\else
\language=\csname l@#1\endcsname
\fi
#2}}
\providecommand{\BIBdecl}{\relax}
\BIBdecl

\bibitem{chagnon_2019_optical}
M.~{Chagnon}, ``Optical communications for short reach,'' \emph{J. Lightw.
  Technol.}, vol.~37, no.~8, pp. 1779--1797, 2019.

\bibitem{zhong_2018_digital}
K.~{Zhong}, X.~{Zhou}, J.~{Huo}, C.~{Yu}, C.~{Lu}, and A.~P.~T. {Lau},
  ``Digital signal processing for short-reach optical communications: A review
  of current technologies and future trends,'' \emph{J. Lightw. Technol.},
  vol.~36, no.~2, pp. 377--400, 2018.

\bibitem{wiegart_2020_probabilistically}
T.~{Wiegart}, F.~{Da Ros}, M.~P. {Yankov}, F.~R. {Steiner}, S.~{Gaiarin}, and
  R.~D. {Wesel}, ``Probabilistically shaped 4-{PAM} for short-reach {IM/DD}
  links with a peak power constraint,'' \emph{J. Lightw. Technol.}, 2020.

\bibitem{ghiasi_2012_investigation}
A.~Ghiasi, Z.~Wang, and V.~Telang, ``Investigation of {PAM}-4/6/8 signaling and
  {FEC} for 100 {G}b/s serial transmission,'' \emph{IEEE 802.3bm Task Force},
  2012.

\bibitem{chorchos_2019_pam6}
L.~Chorchos, ``{{PAM}-6 generation using 32-{QAM} constellation},'' in
  \emph{Metro and Data Center Optical Networks and Short-Reach Links II}, vol.
  10946.\hskip 1em plus 0.5em minus 0.4em\relax International Society for
  Optics and Photonics, 2019.

\bibitem{boecherer2015pas}
G.~{Böcherer}, F.~{Steiner}, and P.~{Schulte}, ``Bandwidth efficient and
  rate-matched low-density parity-check coded modulation,'' \emph{IEEE Trans.
  Commun.}, vol.~63, no.~12, pp. 4651--4665, 2015.

\bibitem{bocherer_probabilistic_2019-1}
G.~B{\"o}cherer, P.~Schulte, and F.~Steiner, ``Probabilistic shaping and
  forward error correction for fiber-optic communication systems,'' \emph{J.
  Lightw. Technol.}, vol.~37, no.~2, pp. 230--244, Jan. 2019.

\bibitem{boecherer2018principles}
\BIBentryALTinterwordspacing
G.~B{\"o}cherer, \emph{Principles of Coded Modulation}.\hskip 1em plus 0.5em
  minus 0.4em\relax Technische Universit{\"a}t M{\"u}nchen, 2018. [Online].
  Available: \url{http://www.georg-boecherer.de/bocherer2018principles.pdf}
\BIBentrySTDinterwordspacing

\bibitem{wesel2001constellation}
R.~D. Wesel, X.~Liu, J.~M. Cioffi, and C.~Komninakis, ``Constellation labeling
  for linear encoders,'' \emph{IEEE Trans. on Inf. Theory}, vol.~47, no.~6, pp.
  2417--2431, 2001.

\bibitem{buchali2018flexible}
F.~Buchali, Q.~Hu, M.~Chagnon, K.~Schuh, L.~Schmalen, and S.~Makovejs,
  ``Flexible transmission enabled by novel m2-qam formats with record
  distance-spectral efficiency tuneability,'' in \emph{Optical Fiber
  Communication Conference}.\hskip 1em plus 0.5em minus 0.4em\relax Optical
  Society of America, 2018, pp. W2A--54.

\bibitem{schulte2015constant}
P.~Schulte and G.~B{\"o}cherer, ``Constant composition distribution matching,''
  \emph{IEEE Trans. Inf. Theory}, vol.~62, no.~1, pp. 430--434, 2015.

\bibitem{frank1953pulse}
F.~Gray, ``Pulse code communication,'' Mar.~17 1953, {US} Patent 2,632,058.

\bibitem{boecherer_2014_labeling}
G.~{Böcherer}, ``Labeling non-square {QAM} constellations for one-dimensional
  bit-metric decoding,'' \emph{IEEE Commun. Lett.}, vol.~18, no.~9, pp.
  1515--1518, 2014.

\bibitem{3gpp-ts-38.212-v15.0.0-17-12a}
``{{3GPP TS}} 38.212 {{V15}}.0.0: {{Multiplexing}} and channel coding,'' Dec.
  2017.

\bibitem{sheikh2017achievable}
A.~Sheikh, A.~G. i~Amat, and G.~Liva, ``Achievable information rates for coded
  modulation with hard decision decoding for coherent fiber-optic systems,''
  \emph{J. Lightw. Technol.}, vol.~35, no.~23, pp. 5069--5078, 2017.

\end{thebibliography}

\end{document}